\documentclass[letter,longauth]{aa}  

\usepackage{graphicx}
\usepackage{txfonts}
\usepackage[]{hyperref}

\defcitealias{2020Sci...369.1347}{M20}

\hypersetup{
    colorlinks=true,
    citecolor=blue,
    linkcolor=blue,
    filecolor=magenta,      
    urlcolor=cyan,
}
\titlerunning{A persistent excess of galaxy-galaxy strong lenses}
\authorrunning{Meneghetti et al.}

\begin{document}

   \title{A persistent excess of galaxy-galaxy strong lensing observed in galaxy clusters}

   \author{Massimo Meneghetti\inst{\ref{oas},\ref{infnbo},\ref{icsc}}\thanks{\email{massimo.meneghetti@inaf.it}} 
   \and
   Weiguang Cui\inst{\ref{uam},\ref{ciaff},\ref{uniedi}}
   \and
   Elena Rasia\inst{\ref{oats},\ref{ifpu}}
   \and
   Gustavo Yepes\inst{\ref{uam},\ref{ciaff}}
   \and
   Ana Acebron\inst{\ref{unimi}}
   \and
   Giuseppe Angora\inst{\ref{unife},\ref{oana}}
   \and
   Pietro Bergamini\inst{\ref{unimi},\ref{oas}}
   \and
   Stefano Borgani\inst{\ref{units},\ref{oats},\ref{ifpu},\ref{infnts},\ref{icsc}}
   \and
   Francesco Calura\inst{\ref{oas}}
   \and
   Giulia Despali\inst{\ref{unibo}, \ref{oas}}
   \and
   Carlo Giocoli\inst{\ref{oas},\ref{infnbo}}
   \and
   Giovanni Granata\inst{\ref{unimi},\ref{oami}}
   \and
   Claudio Grillo\inst{\ref{unimi},\ref{oami}}
   \and
   Alexander Knebe\inst{\ref{uam},\ref{ciaff},\ref{icras}}
   \and
   Andrea V. Macci\`o\inst{\ref{abu},\ref{abu2},\ref{mpia}}
   \and
   Amata Mercurio\inst{\ref{unisa},\ref{oana}}
   \and
   Lauro Moscardini\inst{\ref{unibo},\ref{oas},\ref{infnbo}} 
   \and
   Priyamvada  Natarajan\inst{\ref{yaleastro},\ref{yalephysics}}
   \and
   Antonio Ragagnin\inst{\ref{oas},\ref{ifpu}}
   \and
   Piero Rosati\inst{\ref{unife}}
   \and
   Eros Vanzella\inst{\ref{oas}}
         }
   \institute{
    INAF-Osservatorio di Astrofisica e Scienza dello Spazio di Bologna, Via Piero Gobetti 93/3, I-40129 Bologna, Italy, \label{oas}
    \and
    INFN-Sezione di Bologna, Viale Berti Pichat 6/2, I-40127 Bologna, Italy\label{infnbo}
    \and
     Departamento de F\'isica Te\'orica, M\'odulo 8, Facultad de Ciencias, Universidad Aut\'onoma de Madrid, E-28049 Madrid, Spain \label{uam}
     \and
     Centro de Investigaci\'on Avanzado en F\'isica Fundamental (CIAFF), Facultad de Ciencias, Universidad Aut\'onoma de Madrid, E-28049 Madrid, Spain \label{ciaff}
     \and
     Institute for Astronomy, University of Edinburgh, Royal Observatory, Edinburgh EH9 3HJ, United Kingdom \label{uniedi}
     \and
    INAF – Osservatorio Astronomico di Trieste, via Tiepolo 11, I-34131 Trieste, Italy \label{oats}
    \and
     IFPU – Institute for Fundamental Physics of the Universe, via Beirut 2, 34151, Trieste, Italy \label{ifpu}
     \and
     INFN, Istituto Nazionale di Fisica Nucleare, Via Valerio 2, I-34127, Trieste, Italy \label{infnts}
     \and
     Dipartimento di Fisica, Università degli Studi di Milano, Via Celoria 16, I-20133 Milano, Italy\label{unimi}
     \and
     Dipartimento di Fisica e Scienze della Terra, Università degli Studi di Ferrara, Via Saragat 1, I-44122 Ferrara, Italy\label{unife}
    \and
    INAF-Osservatorio Astronomico di Capodimonte, Via Moiariello 16, 80131 Napoli, Italy\label{oana}
    \and 
    Astronomy Unit, Department of Physics, University of Trieste, via Tiepolo 11, I-34131 Trieste, Italy \label{units}
    \and
    Dipartimento di Fisica e Astronomia "Augusto Righi", Alma Mater Studiorum Università di Bologna, via Gobetti 93/2, I-40129 Bologna, Italy\label{unibo}
    \and
    INAF - IASF Milano, via A. Corti 12, I-20133 Milano, Italy\label{oami}
    \and
    International Centre for Radio Astronomy Research, University of Western Australia, 35 Stirling Highway, Crawley, Western Australia 6009, Australia\label{icras}
     \and
    New York University Abu Dhabi, PO Box 129188 Abu Dhabi, United Arab Emirates\label{abu}
    \and
    Center for Astro, Particle and Planetary Physics (CAP$^3$), New York University Abu Dhabi, United Arab Emirates\label{abu2}
    \and
    Max-Planck-Institut f\"ur Astronomie, K\"onigstuhl 17, D-69117 Heidelberg, Germany\label{mpia}
    \and
    Dipartimento di Fisica “E.R. Caianiello”, Universit`a Degli Studi di Salerno, Via Giovanni Paolo II, I–84084 Fisciano (SA), Italy \label{unisa}
    \and 
    Department of Astronomy, Yale University, New Haven, CT 06511, USA\label{yaleastro}
    \and
    Department of Physics, Yale University, New Haven, CT 06520, USA\label{yalephysics}
    \and
    ICSC - Italian Research Center on High Performance Computing, Big Data and Quantum Computing, Italy\label{icsc}
}

   \date{Received May 23, 2023; accepted September 9, 2023}

 
  \abstract
   {Previous studies have revealed that the estimated probability of galaxy-galaxy strong lensing in observed galaxy clusters exceeds the expectations from the $\Lambda$ Cold Dark Matter cosmological model by one order of magnitude.}
   {We aim to understand the origin of this excess by analyzing a larger set of simulated galaxy clusters, and investigating how the theoretical expectations vary under different adopted prescriptions and numerical implementations of star formation and feedback in simulations.}
   {We performed a ray-tracing analysis of 324 galaxy clusters from the {\sc Three Hundred} project, comparing the {\sc Gadget-X} and {\sc Gizmo-Simba} runs. These simulations, which start from the same initial conditions, were performed with different implementations of hydrodynamics and galaxy formation models tailored to match different observational properties of the intracluster medium and cluster galaxies.}
   {We find that galaxies in the {\sc Gizmo-Simba} simulations develop denser stellar cores than their {\sc Gadget-X} counterparts. Consequently, their probability for galaxy-galaxy strong lensing is higher by a factor of $\sim 3$. This increment is still insufficient to fill the gap with observations as a discrepancy by a factor $\sim 4$ still persists. In addition, we find that several simulated galaxies have Einstein radii that are too large compared to observations.}
   {We conclude that a persistent excess of galaxy-galaxy strong lensing exists in observed galaxy clusters. The origin of this discrepancy with theoretical predictions is still unexplained in the framework of the cosmological hydrodynamical simulations. This might signal a hitherto unknown issue with either the simulation methods or our assumptions regarding the standard cosmological model.}


   \maketitle
%

\section{Introduction}

In the Cold Dark Matter (CDM) model, dark matter halos form through a hierarchical process in which smaller halos form first and subsequently merge to form larger ones. Consequently, dark matter halos in this framework should contain a hierarchy of substructures (or subhalos) \citep{2008MNRAS.386.2135G,2010MNRAS.404..502G}. Furthermore, numerical simulations show that these subhalos should have cuspy density profiles whose shape is well described by a double power law with a logarithmic slope shallower near the center than in the outskirts. Their concentrations and masses should be anticorrelated  \citep[e.g.,][]{1997ApJ...490..493N,2014ApJ...797...34M,2020Natur.585...39W}. These predictions of the CDM model  can be tested against observational data on a wide range of scales, ranging from dwarf galaxies to galaxy clusters, employing various techniques. These include counting the satellites in the neighborhood of the Milky Way \citep{1999ApJ...522...82K,1999ApJ...524L..19M}; detecting signatures of subhalos as perturbations and gaps in stellar streams \citep[e.g.,][]{2019ApJ...880...38B}; mapping the total mass distribution in galaxy clusters utilizing the galaxy--subhalo connection \citep[e.g.,][]{1997MNRAS.287..833N,2015ApJ...800...38G,2017MNRAS.468.1962N} or  the  gravitational imaging technique \citep[e.g.,][]{2005MNRAS.363.1136K,2012Natur.481..341V,2016ApJ...823...37H,2022MNRAS.510.2480D}; and examining flux-ratio anomalies in strong lensing galaxies \citep[e.g.,][]{1998MNRAS.295..587M,2002ApJ...572...25D}.

In this letter we focus on a more recent test of the small-scale structure of galaxy clusters proposed by \cite{2020Sci...369.1347M} (hereafter ME20), who used the galaxy-galaxy strong lensing (GGSL) probability as a metric to measure the compactness and abundance of dark-matter subhalos. A few GGSL events per massive galaxy cluster have been detected in deep observations of cluster cores carried out by the Hubble Space Telescope (HST). The observed ability of galaxy clusters to produce such events requires that their galaxy members have total mass distributions compact enough to become critical for strong lensing. Comparing high-fidelity mass reconstructions of strong lensing clusters with results from numerical hydrodynamical simulations, ME20 found that the observed probability of GGSL is higher than expected in the framework of the $\Lambda$ Cold Dark Matter ($\Lambda$CDM) model by nearly one order of magnitude. Some inconsistency also emerged regarding the spatial distribution of cluster galaxies in simulated clusters compared to observations, where observed cluster members appear more centrally concentrated than their simulated counterparts \citep[see also][]{2017MNRAS.468.1962N}.

\cite{2021MNRAS.505.1458B} and \cite{2021MNRAS.504L...7R} suggested that numerical hydrodynamical simulations with higher mass and spatial resolution, implementing less efficient energy feedback from active galactic nuclei (AGN) compared to the simulations used by ME20, may resolve the above-mentioned discrepancy between theoretical predictions of the GGSL probability and observations. However, \cite{ragagnin22} and \cite{2022A&A...668A.188M} (hereafter ME22) used higher-resolution re-simulations of the same clusters employed by ME20, re-calculated the GGSL probability, and found that the results are insensitive to increased resolution. Moreover, they both showed that reducing the efficiency of AGN feedback to suppress star formation actually enhances the GGSL probability in the simulated clusters, but at the price of producing unrealistic overly massive galaxies that do not match the observed cluster member galaxy population. 

To further confirm the existence of such a significant discrepancy between galaxy clusters simulated in the framework of the CDM model and observations would signal either an unidentified problem with simulation methods or standard cosmological assumptions. Here we carry out the same analysis as ME20 using two larger sets of simulated galaxy clusters from {\sc The Three Hundred} project \citep{2018MNRAS.480.2898C}. The number of lens planes used in the analyses is $\sim 35$ times larger than in ME20, thus significantly increasing the statistical significance of the results. These simulations were obtained using two independent codes, {\sc Gadget} \citep{2005MNRAS.361..776Springel} and {\sc Gizmo} \citep{2015MNRAS.450...53H}. The two codes are based on two completely different hydrodynamical solvers; they also implement very different galaxy formation models tailored to principally reproduce the observed properties of the intracluster medium (ICM) and its scaling relations \citep[][ similar to ME20]{2015ApJ...813L..17R} and the cluster galaxies \citep{2019MNRAS.486.2827D}, respectively. It is of utmost importance to investigate whether galaxy formation models that successfully match the cluster members' stellar properties can confirm the discrepancy found by ME20.   
This   letter is organized as follows. In Sects.~\ref{sect:obsdataset} and \ref{sect:simdataset} we describe the observational and simulated datasets and summarize the method used to measure the GGSL probability. In Sect.~\ref{sect:results} we discuss the results of the comparison between the simulations and observations. Finally, we report our conclusions in Sect.~\ref{sect:conclusions}.

\section{Observational dataset}
\label{sect:obsdataset}
The observational dataset used in this work comprises five galaxy clusters, namely Abell 2744\footnote{For Abell 2744, which has a complex mass distribution characterized by several subclusters, we only consider  a region of $100''\times 100''$ centered on the main cluster, identified by the two brightest cluster galaxies \citep[see BCG1 and BCG2 in][]{2023arXiv230310210B}.} \citep[$z=0.3072$,][]{1998MNRAS.296..392A,2010MNRAS.407...83E}, MACSJ0416.1-2403 \citep[$z=0.397$,][]{2016ApJS..224...33B}, Abell S1063  \citep[$z=0.3457$,][]{2013A&A...559L...9B}, MACSJ1206.2-0847 \citep[$z=0.439$,][]{2013A&A...558A...1B}, and PSZ1-G311.65-18.48 \cite[$z=0.443$,][]{2016A&A...590L...4D}. As presented in previous works, we reconstructed the total mass distribution of these clusters using hundreds of spectroscopically confirmed multiple images of strongly lensed distant galaxies. For these reconstructions we employed a novel technique that combines parametric lens modeling with {\sc Lenstool} \citep[e.g.,][]{1996ApJ...471..643K,2007NJPh....9..447J,2011A&ARv..19...47K,2017MNRAS.472.3177M} with velocity-dispersion priors for the cluster members measured from spectroscopy with the Multi Unit Spectroscopic Explorer (MUSE)  \citep{2019A&A...631A.130B,2021A&A...645A.140B,2021A&A...655A..81P,2022arXiv220709416B,2022arXiv220814020B,2023arXiv230310210B}. Due to the folding in of additional stellar kinematics data, the resulting lens models from this approach are particularly robust on the scales of cluster galaxies relevant for this work.

For each cluster we compute the probability for GGSL following the method proposed by ME20 and including the modifications introduced by ME22.\footnote{The modifications introduced by ME22 corrected the previous underestimation of the GGSL probabilities in ME20.} In short, the procedure involves the following steps. First, we use the lens model to compute the tangential critical lines for a given source redshift. The tangential critical lines, $\vec\theta_c$, correspond to the zero-level contours of the map of the Jacobian tangential eigenvalue, 
\begin{equation}
\lambda_t = 1-\kappa(\vec\theta)-\gamma(\vec\theta) \;,
\end{equation}
where $\kappa(\vec\theta)$ and $\gamma(\vec\theta)$ are the lens convergence and shear \citep[e.g.,][]{Meneghetti2021} obtained from the lens model.

We identify the critical lines connected to the cluster galaxies. We dub these critical lines secondary to distinguish them from the primary critical lines corresponding to the large-scale cluster dark matter halos. Using the deflection angles, $\hat{\vec\alpha}(\vec\theta)$, computed with the cluster lens model, we map these critical lines onto the source plane, obtaining the caustics $\vec\beta_c$:
\begin{equation}
\vec\beta_c=\vec\theta_c-\frac{D_{LS}}{D_S}\hat{\vec\alpha}(\vec\theta_c) \;.
\end{equation}
In the previous equation, $D_{LS}$ and $D_{S}$ are the angular diameter distances between the lens and the source planes and between the observer and the source plane, respectively. 

For each caustic, we compute the enclosed area. Summing the areas $A_{cau,i}$ of all $n_{cau}$ secondary caustics, we obtain the GGSL cross section,
\begin{equation}
\sigma_{\rm GGSL}(z_S) = \sum_i^{n_{cau}} A_{cau,i}(z_S) \;.
\end{equation}

Finally, we divide the GGSL cross section by the area sampled by the cluster mass reconstruction (i.e., the area within which cluster galaxies have been identified and included in the lens model) mapped onto the source plane, $A_{s}(z_S)$, and we obtain the GGSL probability:
\begin{equation}
P_{\rm GGSL}(z_S) = \frac{\sigma_{GGSL}(z_S)}{A_s(z_S)}\;. 
\end{equation}
The same procedure is repeated for several source redshifts, namely $z_S=[1,3,6]$.

The size of the critical lines is quantified by means of their equivalent Einstein radius, defined as
\begin{equation}
    \theta_E = \sqrt{\frac{A_{crit}}{\pi}} \;,
\end{equation}
where $A_{crit}$ is the area enclosed by the critical line. The same definition of Einstein radius applies also to the primary critical lines. In the following sections we   use  $\theta_{E,p}$ to indicate the Einstein radius of the cluster's primary critical line.

\begin{figure*}
   \centering
   \includegraphics[width=1.0\linewidth]{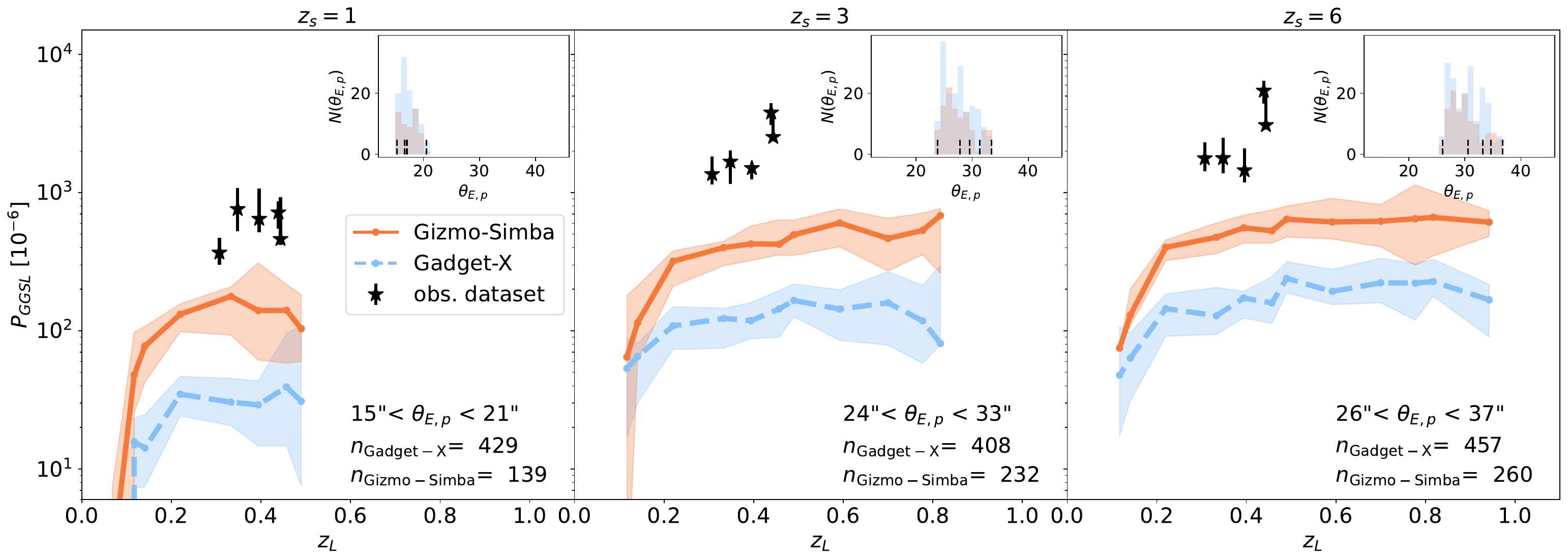}
      \caption{Median GGSL probability as a function of   lens redshift for galaxy clusters with primary Einstein radii $\theta_{E,p}$ in the range of the observational dataset. From left to right, the three panels refer to source redshifts $z_S=1$, $z_S=3$, and $z_S=6$. The results for the {\sc Gizmo-Simba} and {\sc Gadget-X} runs are shown as solid orange and dashed blue lines. The colored bands show the $99.9\%$ confidence intervals for each dataset. The $\theta_{E,p}$ selection limits are reported in each panel and correspond to the range between the minimum and maximum of the primary Einstein radii of the observed clusters for a given source redshift. The number of cluster projections satisfying the selection criteria in the two simulation datasets ($n_{\rm Gizmo-Simba}$ and $n_{\rm Gadget-X}$) are also reported in each panel. The black stars show the GGSL probability of the clusters in the observational dataset with associated $99.9\%$ confidence limits derived from the posterior distributions of the lens model parameters. The insets in the upper right corners show the distributions of primary Einstein radii in the selected simulation samples, limited to the redshift range $0.3 < z_L < 0.45$. The vertical black dashed lines show the measured values in the observational dataset.}
\label{fig:GGSLthetaE}
\end{figure*}

\section{Simulation dataset}
\label{sect:simdataset}
We repeated the above-mentioned procedure to compute the GGSL probabilities for a large set of simulated galaxy clusters from {\sc The Three Hundred} project. This work compares two sets of these simulations generated from the same initial conditions, namely the {\sc Gadget-X} and the {\sc Gizmo-Simba} runs. Both simulation sets comprise full physics hydrodynamical re-simulations of 
spherical regions of radius $15 \;h^{-1}$ Mpc centered on the 
324 most massive galaxy clusters drawn from the 1 Gpc MultiDark N-body simulation \citep[MDPL2;][]{1996ApJ...466...13K}. In addition to implementing different galaxy formation models, the two codes also differ in the hydrodynamical solver adopted.  {\sc Gadget-X}  is based on the implementation in the original {\sc Gadget} code \citep{2005MNRAS.361..776Springel} of an improved formulation of smoothed particle hydrodynamics (SPH, \citealt{2016MNRAS.455.2110B}), which overcomes most of the limitations of the standard SPH. In {\sc Gizmo} the  hydrodynamical forces are computed instead by resorting to a Godunov scheme to solve the Riemann problem between each pair of gas particles \citep{2015MNRAS.450...53H}.  Specifically, the {\sc Gadget-X} simulations implement the galaxy formation model outlined in \cite{2015ApJ...813L..17R}, while the {\sc Gizmo-Simba} simulations adopt a model from the {\sc Simba} simulation described in \cite{2019MNRAS.486.2827D}, and adapted to the resolution of {\sc The Three Hundred} simulations. 

One of the most interesting characteristics of the {\sc Gizmo-Simba} simulations is that they are calibrated to reproduce several stellar properties of observed cluster galaxies, namely their stellar mass function and colors,  and the mass and age of the brightest cluster galaxies (BCGs) \citep{2022MNRAS.514..977C}. In contrast, the {\sc Gadget-X} simulations are tuned to successfully reproduce the observed gas properties and scaling relations in clusters more specifically.
The code used for the {\sc Gadget-X} simulations is extremely similar to that used for the simulations considered in ME20.

The MDPL2 assumes the best-fit cosmological model from \cite{2016A&A...594A..13P}. 
{The mass resolution in the resimulated regions amounts to $1.27\times 10^{9}\;h^{-1}$ M$_\odot$ and $2.36\times 10^{8}\;h^{-1}$ M$_\odot$ per dark matter and gas particles, respectively. The softening length is $5\;h^{-1} $ kpc.
\cite{2022MNRAS.514..977C} provide a detailed description of the {\sc Gadget-X} and {\sc Gizmo-Simba} runs of the {\sc Three Hundred} project and an in-depth discussion of the differences between the galaxy formation models that they implement (see, e.g., their Table 2).}

The lensing analysis of {\sc The Three Hundred} simulations consists of several data products, whose description can be found in Meneghetti et al. (in prep.) \citep[see also][]{2022MNRAS.513.2178H,2023arXiv230200687E}. This analysis uses the deflection-angle, convergence, and shear maps of the 324 clusters obtained by projecting their particles along the three simulation axes. Since the simulation box is uncorrelated with the cluster orientations, these projection directions can be assumed to be random. The maps cover a $200\times 200$ arcsec field of view with $2048\times 2048$ pixels. {Since the size of reconstructed regions in the observational dataset amounts to $\sim 7.12$ sq. arcmin per cluster on average, for computing the $P_{\rm GGSL}$ we restricted our analysis to the central $160\times 160$ arcsec.} 
We used the maps for all clusters in 16 simulation snapshots covering the redshift range $[0.068-1.32]$. Thus, this analysis uses $324\times 16 \times 3= 15552$ lens planes for both the {\sc Gadget-X} and {\sc Gizmo-Simba} runs. Although we can compare these simulations with the observations only within a  much more limited redshift range ($0.31 \lesssim z \lesssim 0.44$), using all these simulated lenses allows us to make a more robust comparison between the two simulation sets. 

We used the maps to compute the secondary critical lines for three source redshifts: $z_S \in [1,3,6]$. We retained those critical lines whose equivalent Einstein radii are $\theta_E >0''.5$. As shown in ME22, assuming this lower limit when comparing the simulations with the observational dataset ensures that the results do not depend on the mass and spatial resolution of the simulations. Then, we mapped the critical lines onto the caustics and computed the GGSL probability as a function of lens and source redshifts. As done by ME22, when computing the GGSL probability in both simulated and observational datasets, we did not account for critical lines with Einstein radii $\theta_E >3''$. Critical lines larger than this limit are produced by groups of galaxies or massive galaxies not observed in the observational dataset, as discussed in Sect.~\ref{sect:thetaE}.

\section{Results}
\label{sect:results}

\subsection{GGSL probability}

In Fig. \ref{fig:GGSLthetaE} we show the probability of GGSL events   derived from the lensing analysis of the simulations. We use solid orange and dashed blue lines to display the results for the {\sc Gizmo-Simba}  and {\sc Gadget-X} datasets, respectively. {The error bands  show the $99.9\%$ confidence intervals of the median. They were computed by bootstrap sampling the simulated data 1000 times. The confidence interval was constructed by taking the percentile interval of the bootstrap distribution.} From left to right, the three panels refer to source redshifts $z_S=1$, $3$, and $6$.
According to the results of the Frontier Fields Lens Modeling Comparison Project \citep{2017MNRAS.472.3177M}, the Einstein radius $\theta_{E,p}$ is one of the cluster properties best constrained by parametric lens inversion methods. In Table~\ref{table:lensprops} we report the values measured for a source redshift $z_S=6$ for all five clusters in the observational dataset. To match observed and simulated clusters, we selected the simulated clusters such that their primary Einstein radii were consistent with those of the observational dataset. For each source redshift evaluated here, we considered only the simulated clusters with $\theta_{E,p}^{min} \leq \theta_{E,p} \leq \theta_{E,p}^{max}$, where $\theta_{E,p}^{min}$ and $\theta_{E,p}^{max}$ are the smallest and largest primary Einstein radii in the observational dataset. The resulting distributions of $\theta_{E,p}$ for clusters selected in the {\sc Gizmo-Simba}  and {\sc Gadget-X} datasets in the redshift range of the observed clusters are shown in the insets in the upper right corner of each panel.

\begin{table}
\caption{Redshifts, primary Einstein radii, and GGSL probabilities for the five clusters in the observational dataset. The reported values refer to source redshift $z_S=6$.}              
\label{table:lensprops}      
\centering                                      
\begin{tabular}{c c c c}          
\hline\hline                        
cluster & $z$ & $\theta_{E,p}$ & $P_{\rm GGSL}$ \\    
 & & [arcsec] & [$10^{-6}$] \\
\hline                                   
   Abell 2744 & $0.3072$ & $26.03_{-0.11}^{+0.29}$ & $1782_{-347}^{+537}$ \\     
   Abell S1063 & $0.3457$ & $36.79_{-0.45}^{+0.34}$ & $1772_{-396}^{+713}$ \\
   MACSJ0416.1-2403 & $0.397$ & $31.31_{-0.35}^{+0.22}$ & $1449_{-263}^{+648}$ \\
   MACSJ1206.2-0847 & $0.439$ & $33.19_{-0.15}^{+0.28}$ & $5456_{-1073}^{+992}$ \\
   PSZ1-G311.65-18.48 & $0.443$ & $34.69_{-0.02}^{+0.69}$ & $ 3074_{-210}^{+2313}$ \\
\hline                                             
\end{tabular}
\end{table}

Independently of the source redshift, we find that clusters in the {\sc Gizmo-Simba} run have GGSL probabilities higher by a factor $\sim 3$ compared to their analogs in the {\sc Gadget-X} simulations. This result also holds outside the redshift range of the observed clusters. The origin of this difference resides in the $H_2$-based star formation model implemented in the {\sc Gizmo-Simba} simulations \citep{2016MNRAS.462.3265D}. A given gas element's star-formation rate (SFR) in this model depends on its $H_2$ molecular fraction. Under this prescription, galaxies develop denser stellar cores at high redshift compared to the {\sc Gadget-X} run, where the SFR is regulated by the gas density and temperature \citep{2023MNRAS.tmp.1468L}. To compensate, stronger feedback is implemented to quench the galaxies to match the observed satellite stellar mass function. Among its successes, this model reproduces the number counts of high-$z$ submillimeter-selected galaxies \citep{2021MNRAS.502..772L}. The drawback is that several properties of the ICM in the inner cluster regions, such as the gas mass density, temperature, and entropy profiles,  are not equally well reproduced \citep{2023MNRAS.tmp.1468L}. In particular, the clusters in the {\sc Gizmo-Simba} run have lower central gas and stellar mass densities compared to the {\sc Gadget-X} simulations. Consequently, these clusters have $\theta_{E,p}$ smaller by $\sim 20\%$ on average. For this reason, as shown in Fig.~\ref{fig:GGSLthetaE}, there are fewer clusters with $\theta_{E,p}$ matching the observational dataset in {\sc Gizmo-Simba} than in the {\sc Gadget-X} simulations. 

Because of their higher central densities, the galaxies in the {\sc Gizmo-Simba} simulation set are more resistant to tidal stripping and can survive as stronger lenses as they fall toward the cluster center. However, despite their higher GGSL probabilities, these simulations still do not match the observations. The black stars indicate the GGSL probabilities of all clusters in the observational dataset. The error bars show the $99.9\%$ confidence limits, derived by randomly sampling the posterior distributions of the lens model parameters. The observed $P_{\rm GGSL}$ exceeds that measured in the {\sc Gizmo-Simba} and in the {\sc Gadget-X} datasets by a factor of $\sim 4$ and $\sim 12$, respectively.

\subsection{Secondary critical lines}
\label{sect:thetaE}
\begin{figure}
   \centering
   \includegraphics[width=1.0\linewidth]{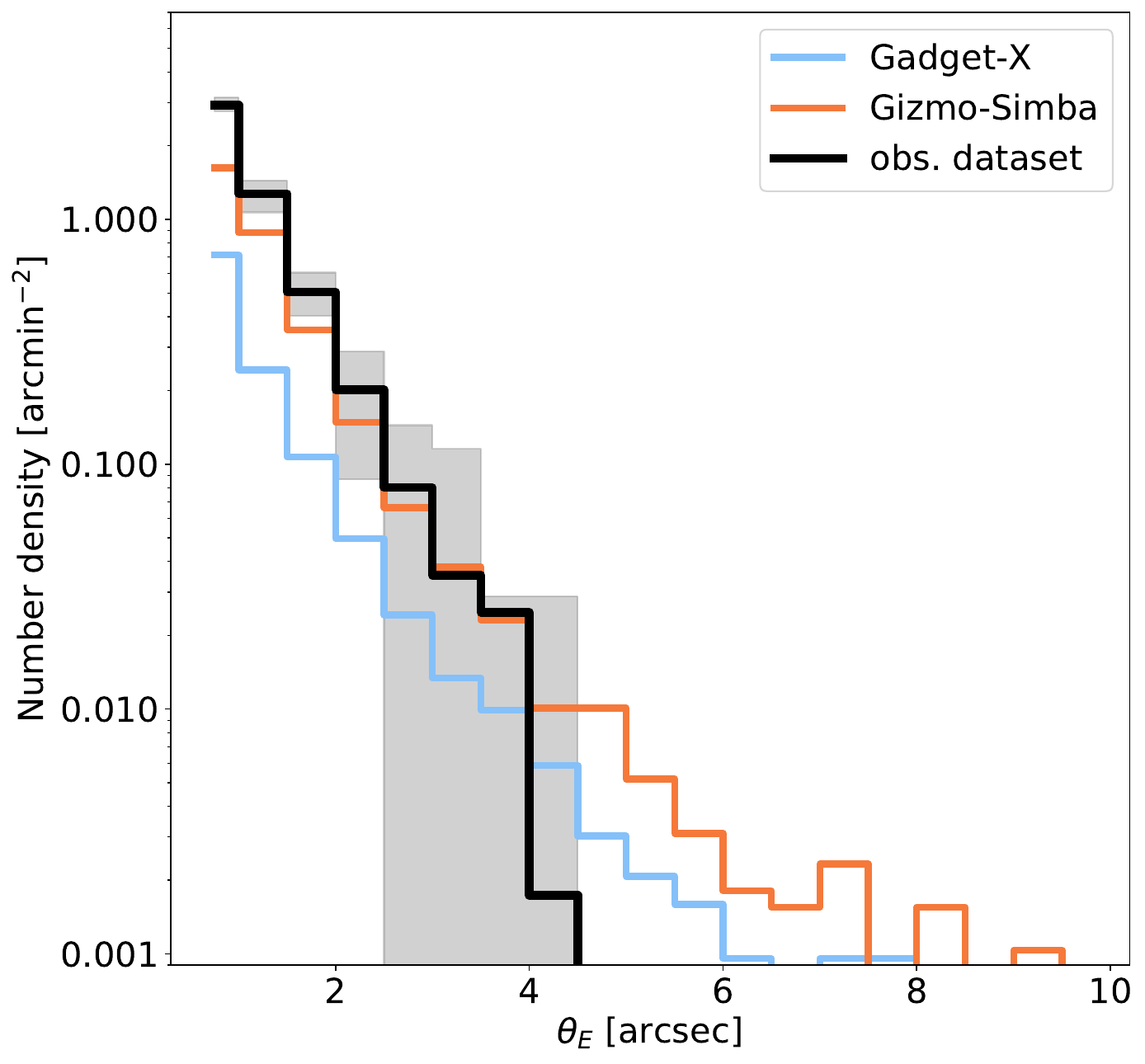}
      \caption{Number density of secondary critical lines (for $z_s=6$) as a function of their equivalent Einstein radius $\theta_E$. The results for the observational dataset are shown with the black histogram. The gray band indicates {the range between the minimum and maximum number density among 100 cluster realizations obtained by randomly sampling the posterior distributions of the lens model parameters}. The light blue and orange histograms show the results for the {\sc Gadget-X} and {\sc Gizmo-Simba} simulation datasets, respectively. For this figure   only clusters with Einstein radii $\theta_{E,p}(z_S=6)>26''$ in the redshift range $0.3 < z_L < 0.45$ were used.}
\label{fig:thetaeprob}
\end{figure}
 
The galaxy formation model implemented in the {\sc Gizmo-Simba} simulations is calibrated to reproduce several observed properties of cluster galaxies. To assess whether this calibration can also reproduce the observed distribution of secondary critical line sizes, we used the snapshots whose redshifts closely match   those of the clusters in the observational dataset, and consider only the galaxy clusters with primary Einstein radii $\theta_{E,p} (z_S=6)>26''$. The number density of secondary critical lines as a function of the size of the Einstein radius is shown in Fig.~\ref{fig:thetaeprob}.  As the black histogram indicates, the secondary critical lines in the observational dataset have Einstein radii (computed assuming $z_S=6$) in the range $0''.5 \lesssim \theta_E \lesssim 4''$. The histogram is the median among 100 realizations per cluster, obtained by randomly sampling the posterior distributions of the {\sc Lenstool} model parameters. The gray band indicates {the range between the minimum and maximum number density among the sampled cluster realizations}. We   verified that the critical lines with $\theta_E \gtrsim 3''$ surround a group of four nearby galaxies in Abell 2744. Therefore, these critical lines are not associated with the production of GGSL events.

The number densities of secondary critical lines in the simulations are generally lower than in the observational dataset, especially noticeable in the {\sc Gadget-X} run. The {\sc Gizmo-Simba} simulations are closer to the observations, but produce an excess of secondary critical lines with large Einstein radii. We also find a similar surplus of secondary critical lines associated with larger Einstein radii in the {\sc Gadget-X} simulations. {The number of galaxy clusters in the observational dataset is small. To estimate the significance of the excess of secondary critical lines noted in the simulations compared to the observations, we generated 1000 samples containing five clusters each by randomly drawing them from the {\sc Gizmo-Simba} dataset. We counted how many of these samples do not contain secondary critical lines with $\theta_E>3''$ and $\theta_E>4''$. They are $1\%$ and $5.4\%$ of the total, respectively.} 

We did not consider the contribution to the GGSL cross sections from galaxies with Einstein radii $\theta_E>3''$ because they are associated with structures not present in the observations. Incorrectly considering their contribution would lead to underestimating the gap between GGSL probabilities in simulated and observed galaxy clusters. In addition, a comparison of quantities needs to be performed on a like-to-like basis. As shown in Fig.~8 of ME22, although they represent only a small percentage of the total, these secondary critical lines would make up a significant fraction (up to $\sim 90\%$) of the GGSL cross section of their host clusters, if accounted for.

The occurrence of large critical lines in the simulated data is related to inefficient AGN energy feedback in the largest galaxies. ME22 showed that some AGN energy feedback schemes, like the one implemented in \cite{2020A&A...642A..37Bassini} (hereafter BA20), whose efficiency at suppressing star formation is relatively low, favor the production of GGSL events. In simulations implementing such AGN feedback schemes, galaxies develop denser stellar cores, making them stronger lenses. The drawback is that these galaxy formation models tend to produce overly massive galaxies that are simply not found in our observational dataset \citep[see, e.g.,][]{ragagnin22}. We find a similar result in the {\sc Gizmo-Simba} run. 

The distribution of Einstein radii in the {\sc Gadget-X} simulations is biased towards large values of $\theta_E$ because many galaxies are not dense enough to produce critical lines individually. However, groups of nearby galaxies can compensate for the low density thanks to their shear, contributing to a more significant fraction of extended critical lines with $\theta_E\gtrsim 3''$ than the {\sc Gizmo-Simba} run. These results agree with those of ME22.

{Although it is unlikely, it is possible that some cluster galaxies have unusually high mass given their luminosities, thus lying outside the scaling relations adopted to model the cluster members with {\sc Lenstool}. In this case, we may incorrectly measure the size of their associated critical lines, especially without strong lensing constraints around them. We note, however, that for all the brightest cluster galaxies, we measure their velocity dispersion spectroscopically, setting strong constraints on their masses.}

\section{Conclusions}
\label{sect:conclusions}

In this letter we followed the method proposed by ME20 and ME22 to measure the GGSL probability in simulations of the {\sc Three Hundred} project. The number of simulated galaxy clusters in this current dataset is $\sim 35$ times larger than was previously used. We analyzed two versions of this set of simulations, performed with independent codes and implementing different star formation models. While ME22 focuses in particular on the impact of various AGN feedback schemes on the GGSL probability, in this work we assessed the sensitivity of the results to other aspects of the galaxy formation models, such as the adopted prescription for star formation. 

Our results can be summarized as follows:
\begin{itemize}
\item The simulated galaxies in the {\sc Gizmo-Simba} set of {\sc The Three Hundred} project have larger cross sections for GGSL, which translate into a GGSL probability higher than in the {\sc Gadget-X} simulation set by a factor of $\sim 3$. This increment in the GGSL probability is due to the $H_2$-based star formation model implemented in the {\sc Gizmo-Simba} run, which produces dense galactic stellar cores at high redshift. Due to their higher central stellar densities, these galaxies are stronger gravitational lenses.
\item Despite the increased production of strong lensing effects, the GGSL probability in the {\sc Gizmo-Simba} run is still below the level measured in our observational dataset by a factor of $\sim 4$. In contrast, the {\sc Gadget-X} simulations fall short of the observations by a factor of $\sim 12$.
\item As in previous results reported in ME22, the Einstein-radii distributions of the secondary critical lines in both the {\sc Gizmo-Simba} and {\sc Gadget-X} runs starkly differ from that of the observational dataset. In the simulations we find a lower spatial density of critical lines with $\theta_E \lesssim 3''$ compared to observations. However,  the simulations produce an excess of secondary critical lines with $\theta_E \gtrsim 3''$ that are simply not found in the observational dataset. 
\end{itemize}

We conclude that current hydrodynamical simulations cannot yet reproduce the GGSL probabilities measured in real galaxy clusters, the demographics of cluster galaxies, and their contributions to the GGSL signal. Interestingly, the galaxies in our observational dataset that contribute to the GGSL probability have masses in the range $\sim 7 \times 10^{10}-10^{12} \;$ M$_\odot$ (see Fig.~11 of ME22 and Fig. S5 of ME20). At lower masses, the secondary critical lines have Einstein radii smaller than the lower limit used in our analysis ($\theta_{E,min}=0''.5$). Galaxies with masses $\gtrsim 10^{12}$ M$_\odot$ do not contribute significantly to the GGSL signal because of the exponential cutoff of the galaxy mass function at high masses.  
Thus, GGSL in galaxy clusters occurs on mass scales where the star formation efficiency reaches its maximum value \citep[or conversely, where the feedback efficiency has minimal impact on star formation; e.g.,][]{2017ARA&A..55..343B,2018ARA&A..56..435W,2020MNRAS.495L..46M}.
Perhaps for this reason we find relatively modest variations of the GGSL probability (by a factor of $\sim 3$) when comparing hydrodynamical simulations that implement different galaxy formation models. 

{In \cite{2023arXiv230906187S}, we further investigated the internal structure of cluster galaxies in the {\sc Three Hundred} project using the maximum circular velocity as a proxy of their compactness}. Our ongoing effort to fully understand the impact of baryon physics on GGSL in galaxy clusters in the context of CDM includes producing simulations at higher resolution with improved galaxy formation models that better match the observed properties of the stellar component of cluster galaxies and the ICM at the same time. On the other hand, our results also suggest that alternative models of dark matter should also be explored in an attempt to explain the observed excess of GGSL in galaxy clusters \citep[e.g.,][]{2021PhRvD.104j3031Y,2022ApJ...926..205C,2022ApJ...937L..30L,2022arXiv221201403M}. 

\begin{acknowledgements}
{We thank the anonymous referee for their constructive comments.} MM was supported by INAF Grant ``The Big-Data era of cluster lensing". We acknowledge financial contributions from PRIN-MIUR 2017WSCC32 and 2020SKSTHZ, INAF ``main-stream'' grants 1.05.01.86.20 and 1.05.01.86.31, by the ICSC National Recovery and Resilience Plan (PNRR)
Project ID CN-00000013 "Italian Research Center on High-Performance
Computing, Big Data and Quantum Computing" funded by MUR Missione 4
Componente 2 Investimento 1.4 - Next Generation EU (NGEU),  by the INFN InDark grant and by ASI n.2018-23-HH.0 grant.
CG and AR are supported by INAF Theory Grant "Illuminating Dark Matter using Weak Lensing by Cluster Satellites".
WC, AK and GY acknowledge Ministerio de  Ciencia e Innovaci\'on (Spain) for partial financial support under research grant PID2021-122603NB-C21. WC is also supported by the STFC AGP Grant ST/V000594/1 and the Atracci\'{o}n de Talento Contract no. 2020-T1/TIC-19882 granted by the Comunidad de Madrid in Spain.
We would also like to thank the Red Espa\~nola de Supercomputaci\'on (RES) for granting us computing resources in  the MareNostrum supercomputer at Barcelona Supercomputing Center, where all the simulations used in this work have been performed.
AK further thanks The Charlatans for the only one I know.
This work was in part performed at the Aspen Center for Physics, which is supported by National Science Foundation grant PHY-2210452. {This material is partially supported by a grant from the Simons Foundation.}
\end{acknowledgements}

%
%
\bibliographystyle{aa} 
\bibliography{bibliography.bib} 

\end{document}